# Artificial Intelligence-Enhanced Couinaud Segmentation for Precision Liver Cancer Therapy

Liang Qiu [a], Wenhao Chi [b], Xiaohan Xing [a], Praveenbalaji Rajendran [c], Mingjie Li [a], Yuming Jiang [d], Oscar Pastor-Serrano [a], Sen Yang [a], Xiyue Wang [a], Yuanfeng Ji [a], Qiang Wen [a,e]

[a] *Department of Radiation Oncology, Stanford University, Stanford, CA 94305 USA*

[b] *Department of Electrical and Computer Engineering, University of Southern California, Los Angeles, CA 90089 USA*

[c] *Massachusetts General Hospital, Harvard Medical School, Boston, MA 02114 USA*

[d] *Department of Radiation Oncology, Wake Forest University, Winston-Salem, NC 27109 USA*

[e] *Department of Radiation Oncology, Shandong Provincial Hospital Affiliated to Shandong First Medical University, Shandong First Medical University, Jinan, 250021, China*

## Abstract

Precision therapy for liver cancer necessitates accurately delineating liver sub-regions to protect healthy tissue while targeting tumors, which is essential for reducing recurrence and improving survival rates. However, the segmentation of hepatic segments, known as Couinaud segmentation, is challenging due to indistinct sub-region boundaries and the need for extensive annotated datasets. This study introduces LiverFormer, a novel Couinaud segmentation model that effectively integrates global context with low-level local features based on a 3D hybrid CNN-Transformer architecture. Additionally, a registration-based data augmentation strategy is equipped to enhance the segmentation performance with limited labeled data. Evaluated on CT images from 123 patients, LiverFormer demonstrated high accuracy and strong concordance with expert annotations across various metrics, allowing for enhanced treatment planning for surgery and radiation therapy. It has great potential to reduces complications and minimizes potential damages to surrounding tissue, leading to improved outcomes for patients undergoing complex liver cancer treatments.

## 1 Introduction

Liver cancer, predominantly in the form of hepatocellular carcinoma (HCC), presents a significant challenge to global health. As one of the most frequently diagnosed cancers and a leading cause of cancer-related deaths worldwide, it underscores the urgent need for advancements in early detection and precise treatment [1, 2]. The escalating incidence of liver cancer is primarily attributed to the heightened prevalence of hepatitis B and C infections, the widespread occurrence of non-alcoholic fatty liver disease, alcohol abuse, and metabolic disorders [3]. Treatment options for hepatic cancer include several surgical approaches such as hepatic resection [4] and liver transplantation [5], alongside ablative therapies like radiofrequency ablation (RFA) and microwave ablation (MWA), which are effective for small, localized tumors [6]. Systemic therapies, such as chemotherapy and targeted agents like sorafenib and lenvatinib, aim to manage tumor growth and improve survival rates in advanced stages [7, 8]. Locoregional treatments, including transarterial chemoembolization (TACE) [9] and radioembolization (TARE) [10], deliver chemotherapy or radiation directly to tumors, serving as valuable options for unresectable tumors or as neoadjuvant therapy. Additionally, radiation therapy plays an integral role, employing advanced techniques such as stereotactic body radiotherapy (SBRT) [11] and proton therapy [12] to precisely target tumors while minimizing damage to healthy liver tissue, crucial for preserving liver function. These modalities collectively enhance treatment efficacy and patient outcomes, highlighting the significance of a comprehensive, multidisciplinary approach to effectively managing liver cancer. Despite these advancements, the prognosis for liver cancer remains poor, with an overall five-year survival rate around 20% [13]. Continued research and advancements in treatment modalities are essential to improve survival rates and quality of life for individuals affected by liver cancer.

The segmentation of hepatic segments is a critical technical process that is deeply integrated into radiology and surgical practices, playing a pivotal role in precision oncology by enhancing the accuracy of treatment planning and monitoring for liver cancer patients. The liver can be divided into eight segments based on vascular and biliary structures, following Couinaud's classification [14]. Precise Couinaud segmentation allows surgeons to optimize tumor removal while preserving sufficient liver function through meticulous surgical planning, which is a crucial for precision surgery in oncology. Additionally, it facilitates the planning of interventional procedures such as RFA and TACE, ensuring that treatments are accurately targeted to tumor-bearing segments. In radiation therapy, the precise segmentation of Couinaud segments is essential for targeting tumor with precision while minimizing radiation exposure to healthy liver tissue and adjacent organs such as the stomach, intestines, and kidneys. This significantly reduces potential side effects and enhances treatment outcomes. Furthermore, segmenting the liver and accurately identifying tumor locations within specific segments aids in monitoring treatment response over time, enabling precise comparisons of imaging studies before, during, and after treatment, providing clinicians with valuable insights into the effectiveness of therapeutic interventions.

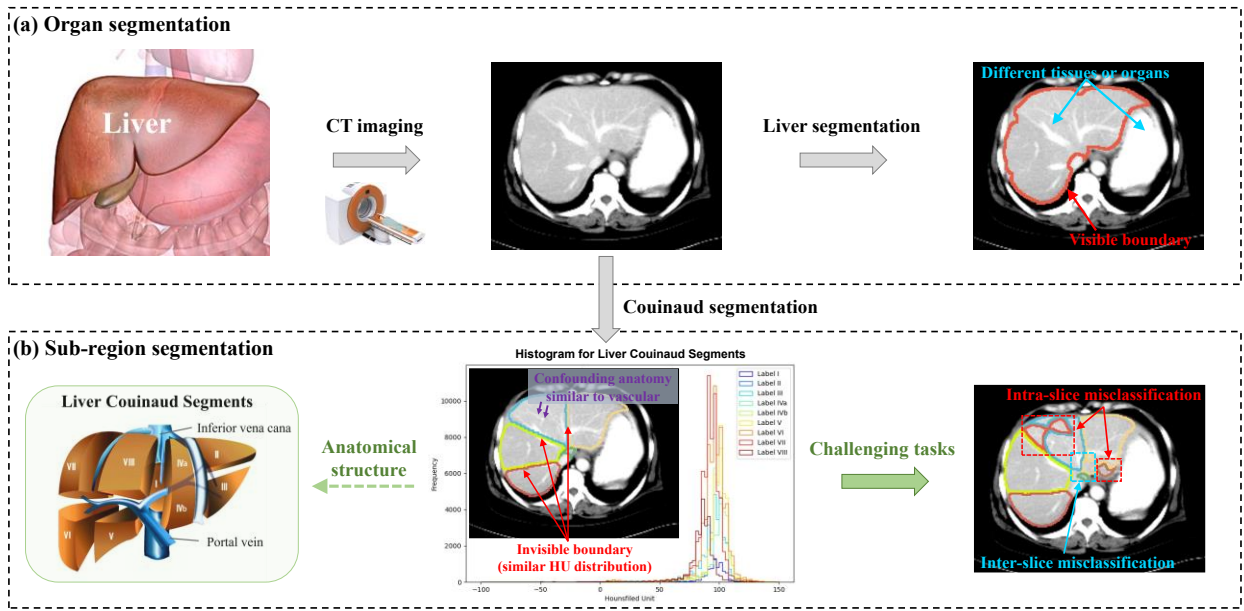

**Fig. 1 Comparison of liver organ-level and sub-region segmentation for liver.** (a) organ segmentation for liver after CT imaging, where the boundary between the liver and other surrounding is relatively clear and visible. (b) Sub-region segmentation of the liver into Couinaud segments based on vascular anatomy. This task is a challenging due to several factors: the similar HU distributions among Couinaud segments, the absence of visible boundary among Couinaud segments, and resemblance of anatomical landmarks to other anatomical features, which often result in intra-slice and inter-slice misclassifications.

Achieving accurate Couinaud segmentation poses several challenges due to the complexity of differentiating the liver's anatomical segments. Unlike whole-organ segmentation [15], where the liver's boundaries are relatively distinct, Couinaud segmentation is hindered by the homogenous distribution of Hounsfield units (HU) among its sub-regions. These boundaries are not readily discernible on imaging and are typically defined by anatomical landmarks that can confounded with other structures in each CT slice (see Fig. 1). Even experienced clinicians may struggle with these nuances, often needing to cross-reference multiple slices to accurately delineate segments. This process is demanding, time-consuming, and subject to variations among different observers and even the same observer at different times. Consequently, there is a substantial need for effective, automated, and standardized tools for Couinaud segmentation in clinical practice, utilizing advanced algorithms to improve both accuracy and efficiency.

Over the past two decades, extensive research has focused on computer-assisted Couinaud segmentation, primarily utilizing traditional machine learning techniques [16-24]. However, these methods often fall short in accuracy and efficiency for clinical applications. The advent of deep learning has revolutionized medical image segmentation, including both whole-organ and sub-region segmentation [25-30]. A prominent model is U-Net, an end-to-end segmentation network that has significantly improved the accuracy and efficiency of Couinaud segmentation in CT images. Nevertheless, the 2D Convolutional Neural Network (CNN) structure processes each slice independently, overlooking crucial information along the z-axis in CT volumes [31, 32]. To address this, a 2.5D deep learning model was developed to enhance 2D context by incorporating adjacent inter-slice features, resulting in more robust segmentation [33]. Despite this, ambiguities within individual slices persist, and the model's performance depends heavily on the precise classification of features within each slice. With the evolution in computing capabilities and GPU resources, 3D deep learning networks have emerged as a promising approach. For instance, a 3D U-Net model utilizing pixel-wise and boundary loss has demonstrated effectiveness in segmenting liver sub-regions from MRI scans [34]. Similarly, A 3D U-Net architecture was employed for the automated delineation of Couinaud segments and future liver remnants on contrast-enhanced portovenous phase CT images [35]. Additionally, deep learning techniques have advanced landmark detection, improving the localization of key points within the Couinaud segments. The ARH-CNet, a cascaded network built on an attentive residual hourglass framework, has been instrumental in accurately identifying bifurcation points within the hepatic vascular network. These landmarks enable precise liver division into functional units, enhancing clinical applications such as tumor localization and post-treatment organ volume assessment [36]. Despite these strides, a common limitation remains: the reliance on supervised learning methods necessitates extensive labeled data, often impractical due to the labor-intensive nature of manual annotation on 3D medical images. Moreover, inherent ambiguities between liver subregions pose challenges that can impact critical clinical tasks, such as tumor localization and post-treatment organ volume assessment.

To address these challenges, we propose LiverFormer, an innovative model tailored for liver Couinaud segmentation, by integrating the robust 3D U-shape architecture and Transformer techniques [37]. The U-shape architecture, exemplified by U-Net, is widely recognized in medical image segmentation for its excellence in local feature extraction through convolutional operations. However, it falls short in modeling long-range dependencies. In contrast, Transformers are inherently equipped with global self-attention mechanisms but may struggle with precise localization. Our method bridges these capabilities by combining the global context from Transformers with high-resolution spatial details from CNN features derived from the U-shape architecture, thereby enhancing the precision of Couinaud segmentation from CT scans. We further introduce a registration-based data augmentation strategy to mitigate the challenge of data scarcity, thereby improving our model's efficacy in real-world clinical settings. Extensive evaluation on our proprietary dataset, alongside comparative analyses with state-of-the-art methodologies, demonstrates LiverFormer's superior accuracy and efficiency. These findings highlight LiverFormer's potential and establish a compelling case for its integration into advanced clinical decision support systems, potentially revolutionizing liver disease management.

## 2 Methods

### 2.1 Dataset ethics

This study was approved by the Research Ethics Committee of Shandong Provincial Hospital and was carried out in compliance with the ethical standards established in the 1964 Helsinki Declaration. Informed consent was obtained from all participants before their inclusion in the study. We retrospectively reviewed CT scans from patients who underwent routine abdominal imaging at Shandong Provincial Hospital from January 2020 and December 2021. For model development, we selected 123 contrast-enhanced CT scans that met our quality criteria. Specifically, images were omitted if they displayed severe liver deformation from advanced cirrhosis or large tumors, if the patient had previously undergone partial hepatectomy or liver transplantation, or if the image quality was degraded by artifacts.

## 2.2 Image acquisition

All participants underwent CT scans using a Philips iCT 128 scanner, following a meticulously designed imaging protocol to ensure high-quality and consistent images for precise analysis. Scans were performed with a tube voltage of 120 kV and a tube current ranging from 300 to 400 mA. The images were acquired with a slice thickness of 3 mm and reconstructed using a matrix size of $512 \times 512$ pixels. The in-plane pixel spacing was $0.8142 \times 0.8142$ mm², providing high spatial resolution for detailed visualization. Scanning was conducted in helical mode to ensure comprehensive anatomical coverage and optimal image quality.

Following a routine non-enhanced scan, contrast-enhanced CT (CECT) scans were performed during the arterial and venous phases. The arterial phase began 35-40 seconds after intravenous injection of iodinated contrast material (Omnipaque, GE Healthcare), while the venous phase started 65–70 seconds post-injection. The contrast agent was given at a dosage of 1.5 mL per kilogram of body weight, with an injection rate of 3.5–4 mL/s, administered through a pump injector. Patients were closely monitored for potential adverse reactions during and after contrast administration to ensure both safety and comfort. The imaging data were subsequently stored in the PACS for further analysis.

## 2.3 Manual segmentation and preprocessing

Manual segmentation of Couinaud segments was performed by two highly experienced radiologists, with substantial backgrounds in abdominal imaging of 7 years and 10 years, respectively (see Fig. 2(a)). The segmentation was conducted utilizing the 3D Slicer platform (Version 4.10) and strictly followed anatomical guidelines established by Couinaud, a standard in hepatic surgery and radiology [14]. Each radiologist independently conducted their manual delineation on a slice-by-slice basis using abdominal window settings (300/−60 Hu), ensuring objectivity and minimizing bias by remaining blinded to the clinicopathological data associated with the subjects. Consistency and accuracy were upheld through consensus discussions to resolve any discrepancies between the radiologists. Additionally, all results underwent double-checking and verification by another experienced radiologist, which validated the segmentation accuracy and assured the reliability of the resulting ground truth dataset. For model development and evaluation, all the images and segmentation labels were linearly aligned and resampled with 1 mm isotropic voxels. They also underwent intensity normalization with range [0, 1], resulting in well-defined images with a resolution of 256 × 256 × 32.

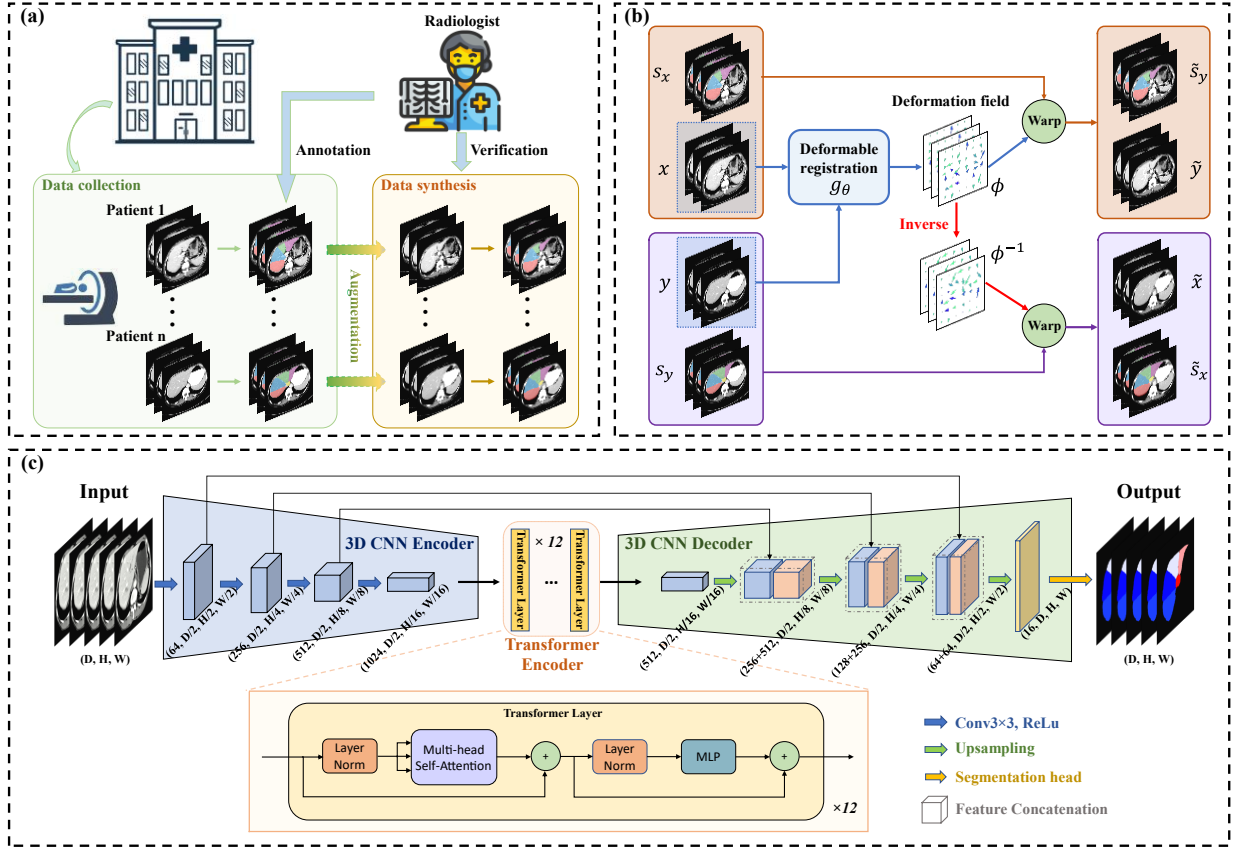

**Fig. 2 Overview of our methodology for liver Couinaud segmentation.** (a) Data collection and preparation. Abdominal CT images of patients are collected from hospitals or clinics, which are annotated by radiologists to establish the manual Couinaud segmentation ground truth. Data augmentation is then performed on these images to synthesize realistic data, further diversifying the dataset for improved model training after verification by radiologists. (b) Data augmentation strategy via deformation registration. The deformable registration tool ANTs is exploited to compute the deformation field $\phi$, which is subsequently applied to original images ($x$, $y$) and labels ($s_x$, $s_y$) to generate synthesized images ($x \circ \phi$, $y \circ \phi^{-1}$) and labels ($s_x \circ \phi$, $s_y \circ \phi^{-1}$). (c) Network design. Our deep learning model, LiverFormer, takes 3D abdominal CT image as input to predict the corresponding predicted Couinaud segments as output. The network consists of three primary components: a 3D CNN-based encoder, a Transformer module and an Upsampling module that integrates a 3D CNN decoder and skip connection operations.

**2.4 Data augmentation**

In liver cancer imaging, precise segmentation of the liver and its internal structures—particularly the Couinaud segments—is essential for accurate diagnosis and effective treatment planning. While supervised deep learning models excel with ample labeled data, manually annotating medical images is labor-intensive and requires significant expertise. Additionally, variations in image acquisition protocols across different machines and institutions—such as differences in resolution, noise levels, and tissue appearance—further complicate the manual annotation process. To address the scarcity of high-quality annotated data, conventional data augmentation techniques—such as random scaling, rotation, and nonlinear deformation—are commonly employed to reduce overfitting and enhance segmentation accuracy by increasing data diversity. However, these methods often fail to capture the intricate anatomical variations inherent in medical images and can be highly sensitive to parameter selection, limiting their ability to provide robust training data. To overcome these limitations, deformable registration can effectively augment the diversity of the original CT datasets by warping the images through a deformation field, and applying complex, non-linear transformations to create variations that mimic different anatomical configurations. This approach effectively augments the diversity of original CT datasets, providing deep learning models with a broader range of training examples that better reflect the variability encountered in clinical practice. In this study, we employ the deformable registration module in Advanced Normalization Tools (ANTs) to generate a diverse set of augmented images (see Fig. 2(b)). Let x, y be any two image volumes with corresponding manual segmentation annotations $s_x$ and $s_y$ within in our 3D liver CT dataset $\Omega \subset \mathbb{R}^3$. We begin by assuming that x and y are affinely aligned during the preprocessing stage, meaning that any misalignment between the volumes is purely nonlinear. We consider a spatial transformation represented by a smooth voxel-wise displacement field, denoted as $u$.

The Symmetric Normalization (SyN) model $g_\theta$ within ANTs is then applied to x and y using a defined set of parameters $\theta$ to compute the resulting deformation $\phi$, shown as follows.

$$g_\theta(x, y; \phi) = Id + u, \tag{1}$$

where $Id$ represents the identity function, and $x \circ \phi$ indicates the process of applying the deformation $\phi$ to $x$. Then we can get the synthesized images $\tilde{x}$ and $\tilde{y}$ along with corresponding segmentation labels $\tilde{s}_x$ and $\tilde{s}_y$ as follows:

$$\tilde{x} = \tau_f(x) = x \circ \phi, \quad \tilde{y} = \tau_b(y) = y \circ \phi^{-1}, \tag{2}$$

$$\tilde{s}_x = \tau_f(s_x) = s_x \circ \phi, \quad \tilde{s}_y = \tau_b(s_y) = s_y \circ \phi^{-1}, \tag{3}$$

where $\tau_f$ is the deformable transformation that transforms $x$ to get the synthesized image $\tilde{x}$ using deformation field $\phi$, and $\tau_b$ is the inverse deformable transformation that transforms $y$ to get the synthesized image $\tilde{y}$ using inverse deformation field $\phi^{-1}$.

This approach effectively enhances the diversity and realism of our limited dataset, mitigates overfitting, and improves model generalization, leading to more accurate and robust segmentation performance.

## 2.5 Model development

In this study, we introduce LiverFormer, an end-to-end network designed to tackle the Couinaud segmentation challenge by leveraging both U-shape architecture and Transformer techniques, as shown in Fig. 2(c). LiverFormer comprises three core components: (a) a 3D CNN encoder that applies 3D convolutions on the entire CT image to capture detailed high-resolution volumetric information; (b) a Transformer module equipped with self-attention mechanisms that further encodes tokenized image patches from the 3D CNN for extracting global feature; and (c) an upsampling module, including 3D CNN decoder and skip connection, that restores segmentation details to align with the input image's original dimensions. These components synergize to boost the accuracy and effectiveness of Couinaud segmentation.

Given a 3D CT image $x \in \mathbb{R}^{D \times H \times W \times C}$ as input, with dimensions D×H×W and C channels, the objective is to generate a pixel-wise label map of the same size (D×H×W) for Couinaud segmentations. We start by employing a 3D CNN architecture, specifically an adapted 3D ResNet [38], which functions to produce the corresponding feature representation. This representation captures finer details through successive convolutional layers operating at various scales. We then reshape the feature map into $N$ smaller, flattened 3D patches $\{x_i^p \in \mathbb{R}^{P^3 \times C} \mid i=1,...,N\}$, each sized $D \times H \times W$. These flattened patches $x^p$ are projected into a hidden $d$-dimensional representation via a learnable linear transformation. To preserve spatial context, positional encodings are incorporated into the patch representations, ensuring the retention of positional information:

$$z_0 = [x_1^p H;\ x_2^p H;\ ...;\ x_N^p H] + H^{pos}, \tag{4}$$

where $H \in \mathbb{R}^{(P^3 \times C) \times d}$ represents the projection of the patch embeddings, and $H^{pos} \in \mathbb{R}^{N \times d}$ indicates the position encodings. Next, the Transformer module encodes embedding features $z_0$ generated by the CNN as its input, aiming to capture global context. Each Transformer layer is composed of a Multihead Self-Attention (MSA) mechanism paired with a Multi-Layer Perceptron (MLP) block. The output of the $l$-th layer is expressed as:

$$z_l' = MSA(LN(z_{l-1})) + z_{l-1}, \tag{5}$$

$$z_l = MLP(LN(z_l')) + z', \tag{6}$$

where $LN(\cdot)$ represents the layer normalization operator and $z_l$ indicates the encoded image representation. This hybrid CNN-Transformer architecture capitalizes on the fine-grained spatial details captured by CNNs and the broad contextual representation offered by Transformers, resulting in a robust encoder for our segmentation task. Drawing inspiration from the U-shaped architecture, the self-attentive features obtained from the Transformers are upsampled and fused with high-resolution CNN features through skip connection. This integration facilitates the decoding of latent representations, ultimately generating the final segmentation output. The upsampling size is carefully adjusted to match the stride of the previous convolution stage, ensuring seamless integration of features across scales. We utilize the multi-class Dice segmentation loss $L_{seg}$ for model training, shown as follows:

$$L_{seg} = 1 - \frac{1}{C}\sum_{c=1}^{C} \frac{2\sum_{i}^{K} p_{i,c} g_{i,c}}{\sum_{i}^{N} p_{i,c}^2 + \sum_{i}^{N} g_{i,c}^2}, \quad (7)$$

where $K$ refers to the total pixel count, $C$ indicates the number of distinct classes, $p_{i,c}$ denotes the predicted probability of pixel $i$ being classified as class $c$, and $g_{i,c}$ is the one-hot encoded ground truth label indicating whether pixel $i$ belongs to class $c$. After training with the augmented dataset, our model demonstrates superior performances in the segmentation task of Couinaud segments based on the design of strong hybrid 3D CNN-Transformer encoder.

## 2.6 Implementation

We used PyTorch to implement the model and trained it end-to-end on an NVIDIA RTX A5000 GPU equipped with 32GB of memory. We employed the Adam optimizer, initializing with a learning rate of 1e-3 and decreasing it by a factor of 0.1 every 50 epochs to promote effective convergence. The training process was guided by a multi-class Dice Loss function over the course of 150 epochs, utilizing a mini-batch size of 1. The dataset was divided into training, validation, and test sets, comprising 87, 18, and 18 patients, respectively. To increase variability within the training dataset, we employed data augmentation techniques using deformable registration with the ANTs tool. This process involved registering selected template images with patient images in the training set to generate more realistic samples. Specifically, we selected one and three images as templates respectively for augmentation in our validation experiments.

## 2.7 Quantitative evaluation

To assess the accuracy and efficiency of Couinaud segmentation, we conducted a quantitative analysis using an internal test dataset comprising 18 cases that were not included from the training stage. The assessment focused on four key metrics: Dice Similarity Coefficient (DSC), Mean Surface Distance (MSD), Hausdorff Distance (HD), and Volume Ratio (RV).

**Dice Coefficient (Dice).** DSC is used to measure the overlap between the predicted segmentation X and the ground truth Y, which is defined as:

$$\text{Dice}(X,Y) = \frac{2|X \cap Y|}{|X|+|Y|} \tag{8}$$

The DSC value ranges from 0 to 1, where a value of 0 represents no overlap while a value of 1 indicates an exact match.

**Mean Surface Distance (MSD).** MSD quantifies the average distance between the surfaces of the predicted segmentation $X$ and the ground truth $Y$. It is computed as:

$$\text{MSD}(X,Y) = \frac{1}{|S_X|+|S_Y|}\left(\sum_{x \in S_X} \min_{y \in S_Y} d(x,y) + \sum_{y \in S_Y} \min_{x \in S_X} d(y,x)\right), \tag{9}$$

where $S_X$ and $S_Y$ are the surface points of $X$ and $Y$, respectively, and $d(x,y)$ is the Euclidean distance between points $x$ and $y$.

**95th Percentile Hausdorff Distance (95HD).** HD quantifies the maximum distance between the surface points of the predicted segmentation $X$ and the ground truth $Y$, providing a worst-case scenario metric. It is defined as:

$$\text{HD}(X,Y) = \max\left(\sup_{x \in S_X} \min_{y \in S_Y} d(x,y), \sup_{y \in S_Y} \min_{x \in S_X} d(y,x)\right), \tag{10}$$

where $S_X$ and $S_Y$ are the surface points of $X$ and $Y$, respectively, and $d(x,y)$ is the Euclidean distance between points $x$ and $y$. The 95HD is a more robust alternative to the HD, emphasizing the distances at the 95th percentile between the boundary points of the predicted segmentation and the corresponding ground truth. This metric is less sensitive to outliers than the maximum HD and provides a more stable and representative measure of boundary errors.

**Volume Ratio (RV)**. RV evaluates the ratio of the volumes of the Couinaud segments from the predicted segmentation $X$ and the ground truth $Y$. It is determined by:

$$RV(X,Y) = \frac{V_X}{V_Y} \tag{11}$$

where $V_X$ and $V_Y$ indicate the volumes of $X$ and $Y$, respectively.

## 3 Experimental Results

### 3.1 Data augmentation using deformable registration

Obtaining accurate Couinaud segmentation is clinically significant but quite challenging, particularly when facing data scarcity. To address this, we explore a feasible data augmentation strategy using deformable registration to synthesize a wide variety of realistic new images, aiming to improve segmentation performance without increasing the annotation burden on doctors. As shown in Fig. 2 (b), we start with two images $x$ and $y$, randomly selected from the dataset, each accompanied by corresponding segmentation labels $s_x$ and $s_y$. Using the deformable registration method integrated within ANTs, we generate a pair of non-linear spatial transformations between these two images, namely the deformation field $\phi$ and its inverse operation $\phi^{-1}$. We generate synthetic labeled samples by modifying the original dataset through various transformations, producing a variety of diverse and authentic labeled instances. In our experiment implementation, we initially selected one patient example at random as the template from 87 training samples out of a total of 123 patients, and conducted the atlas-based deformable registration against the other patients, resulting in an additional 172 synthesized examples. We then incorporate two more templates for comparison, increasing the total number of synthesized examples to 510. Fig. 3 presents two examples of image augmentation for visualization, where the first row shows the original images and labels, and the second row depicts the synthesized images ($x \circ \phi$, $y \circ \phi^{-1}$) and

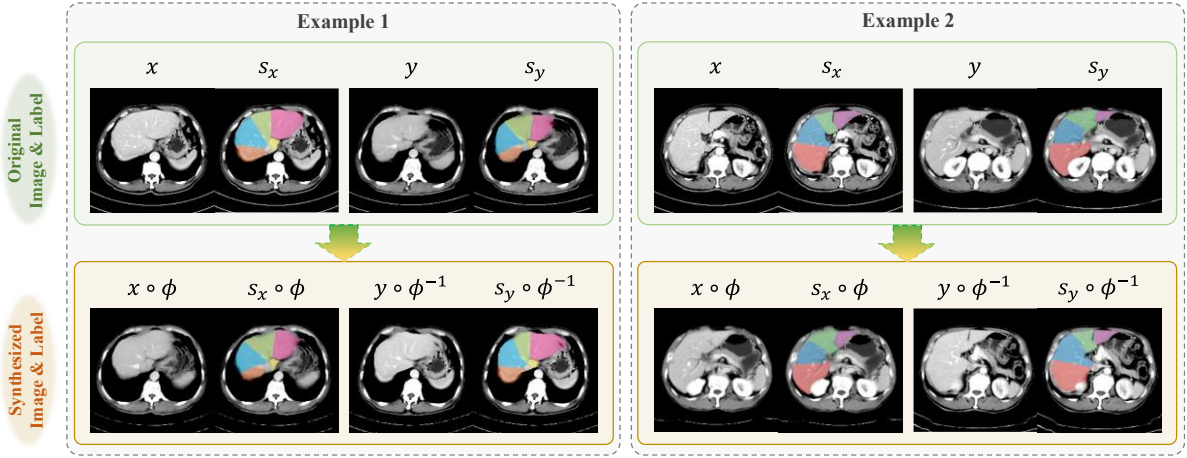

**Fig. 3 Synthesized examples of paired images and labels for dataset augmentation using deformable registration.** The registration field $\phi$ estimated through deformable registration tool ANTs is applied to original images ($x$, $y$) and labels ($s_x$, $s_y$) to generate synthesized images ($x \circ \phi$, $y \circ \phi^{-1}$) and labels($s_x \circ \phi$, $s_y \circ \phi^{-1}$). Our strategy creates highly realistic anatomical variations while preserving the contextual relationships between different anatomical structures with high fidelity, closely mimic real patient variability.

**Table 1 Ablation experiments for data augmentation strategy**

| Method | Dice | MSD (mm) | HD95 (mm) | VR |
|---|---|---|---|---|
| Ours w/o aug | 0.802±0.062 | 2.897±1.488 | 5.823±2.974 | 0.992±0.222 |
| Ours (1 template) | 0.814±0.053 | 2.563±1.920 | 5.558±2.812 | 1.015±0.192 |
| Ours (3 template) | **0.820±0.049** | **2.386±1.123** | **5.225±1.717** | **1.006±0.170** |

labels($s_x \circ \phi$, $s_y \circ \phi^{-1}$). These visualizations offer intuitive insights, demonstrating that our strategy effectively generates anatomical variations that closely resemble real patient variability while preserving contextual relationships between different anatomical structures.

We utilize the expanded dataset, comprising both original and synthesized examples, for training a supervised segmentation model. The results of the ablation study, as shown in Table 1, highlight the efficacy of our data augmentation strategy implemented using our LiverFormer model. Specifically, the comparison between the model trained exclusively on the original dataset (Ours w/o aug) and those trained with augmentation (Ours with 1 template and 3 templates) demonstrates the advantage of incorporating data augmentation. Utilizing a single template leads to performance improvements across all metrics, achieving a higher Dice coefficient (Dice) score of 0.814±0.053 compared to 0.802±0.062 for the non-augmented model. This improvement is even more pronounced with the use of three templates, which results in the highest Dice score (0.820±0.049) and the lowest values for Mean Surface Distance (MSD) and the 95th percentile of Hausdorff Distance (HD95), with MSD dropping to 2.386±1.123 mm and HD95 to 5.225±1.717 mm, reflecting consistent improvements 17.6% and 10.3%, respectively. The Volume Ratio (VR) also shows a favorable outcome for the three-template augmentation strategy, recording 1.006±0.170. These results affirm the substantial benefits of our registration-based data augmentation approach, highlighting its ability to enhance boundary precision and volumetric consistency in Couinaud segmentation tasks.

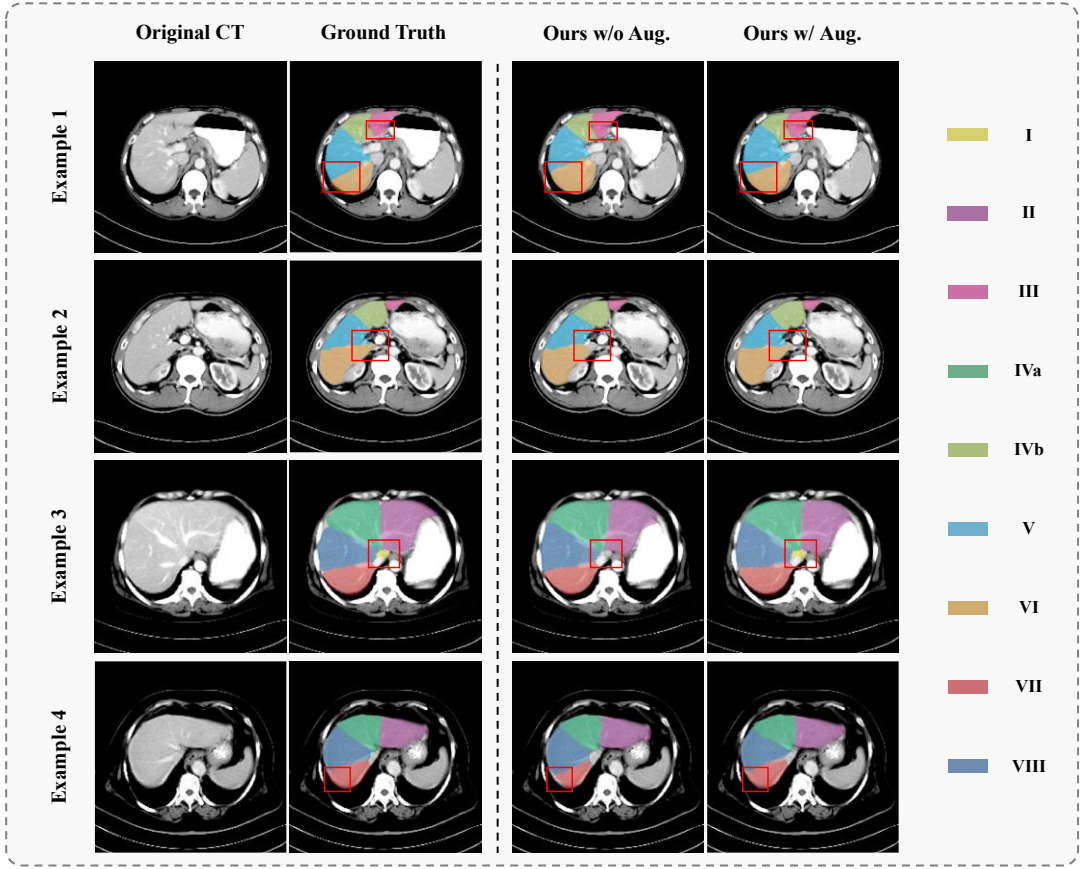

**Fig. 4 Qualitative comparison of Couinaud segmentation outcomes using our data augmentation strategy.** The first two columns present the original CT images and their corresponding ground truth segmentations. The third column presents segmentation results from our model trained without data augmentation, while the fourth column shows results from the model incorporating data augmentation. Each row represents an example from a different patient. The segmented liver sub-regions are represented with masks in different colors. The red boxes highlight the improvements achieved through the augmentation methods, demonstrating the enhanced segmentation accuracy and boundary delineation achieved through our data augmentation approach.

Fig. 4 illustrates a visual comparison of segmentation results for four representative examples from different patients, showcasing the efficacy of our LiverFormer model with data augmentation strategies. Each example displays segmented liver sub-regions using distinct color-coded masks, with red boxes highlighting the improvements achieved through our augmentation methods. Specifically, our augmentation method facilitates the accurate capture of Segment VI and VII for intra-slice segmentation in Examples 2 and 4 respectively, and Segment I for cross-slice segmentation in Example 3. Additionally, it enhances the delineation of interfaces between different hepatic segments, such as the boundary between Segments V and VI, as shown in Example 1. This visual representation complements the quantitative results in Table 1, providing intuitive insights into the model's performance and the tangible benefits of incorporating synthesized data into the training process, leading to more accurate and reliable liver Couinaud segmentation. This visual representation complements the quantitative results in Table 1, providing intuitive insights and underscoring the tangible benefits of incorporating synthesized data into the training process, which leads to more accurate and reliable Couinaud segmentation.

**Table 2 Performance comparison in terms of Dice metric with state-of-the-art approaches**

| Models | I | II | III | IVa | IVb |
|---|---|---|---|---|---|
| 3D U-Net | 0.782±0.081 | 0.744±0.062 | 0.706±0.119 | 0.743±0.088 | 0.737±0.069 |
| 3D V-Net | 0.763±0.102 | 0.760±0.070 | 0.763±0.102 | 0.772±0.077 | 0.759±0.133 |
| TransUNet | 0.800±0.041 | 0.771±0.071 | 0.804±0.085 | 0.764±0.076 | 0.801±0.056 |
| **Ours** | **0.811±0.050** | **0.798±0.051** | **0.841±0.049** | **0.798±0.064** | **0.815±0.048** |
| Models | V | VI | VII | VIII | **Average** |
| 3D U-Net | 0.760±0.063 | 0.795±0.067 | 0.776±0.108 | 0.804±0.070 | 0.761±0.086 |
| 3D V-Net | 0.766±0.119 | 0.823±0.066 | 0.787±0.075 | 0.808±0.049 | 0.778±0.090 |
| TransUNet | 0.778±0.087 | 0.813±0.062 | 0.758±0.060 | 0.797±0.062 | 0.787±0.069 |
| **Ours** | **0.819±0.038** | **0.846±0.042** | **0.814±0.042** | **0.840±0.033** | **0.820±0.049** |

## 3.2 Comparison with state-of-the-art methods

**Quantitative Analysis.** In our experiments, we evaluated our method against the widely recognized 3D U-Net and 3D V-Net models, as well as the advanced medical image segmentation framework, TransUNet. As illustrated in Table 2, our developed model consistently outperformed these methods, achieving the highest Dice scores across all sub-regions (I–VIII) and in the overall average. Specifically, our model's Dice scores range from 0.798 to 0.841, surpassing those of TransUNet and showing a substantial advantage over both 3D U-Net and 3D V-Net in most cases. Additionally, our model attained an impressive average Dice score of 0.820, outperforming the averages of 0.787, 0.778, and 0.761 obtained by TransUNet, 3D V-Net, and 3D U-Net, respectively. Notably, even without data augmentation, our approach maintains an average Dice score of 0.802 (see Table 1), underscoring its clear superiority over other methods. This emphasizes the strength of our hybrid 3D CNN-Transformer architecture, which fuses detailed spatial features from CNNs with the broader contextual insights provided by Transformers, greatly improving the precision and accuracy in Couinaud segmentation tasks.

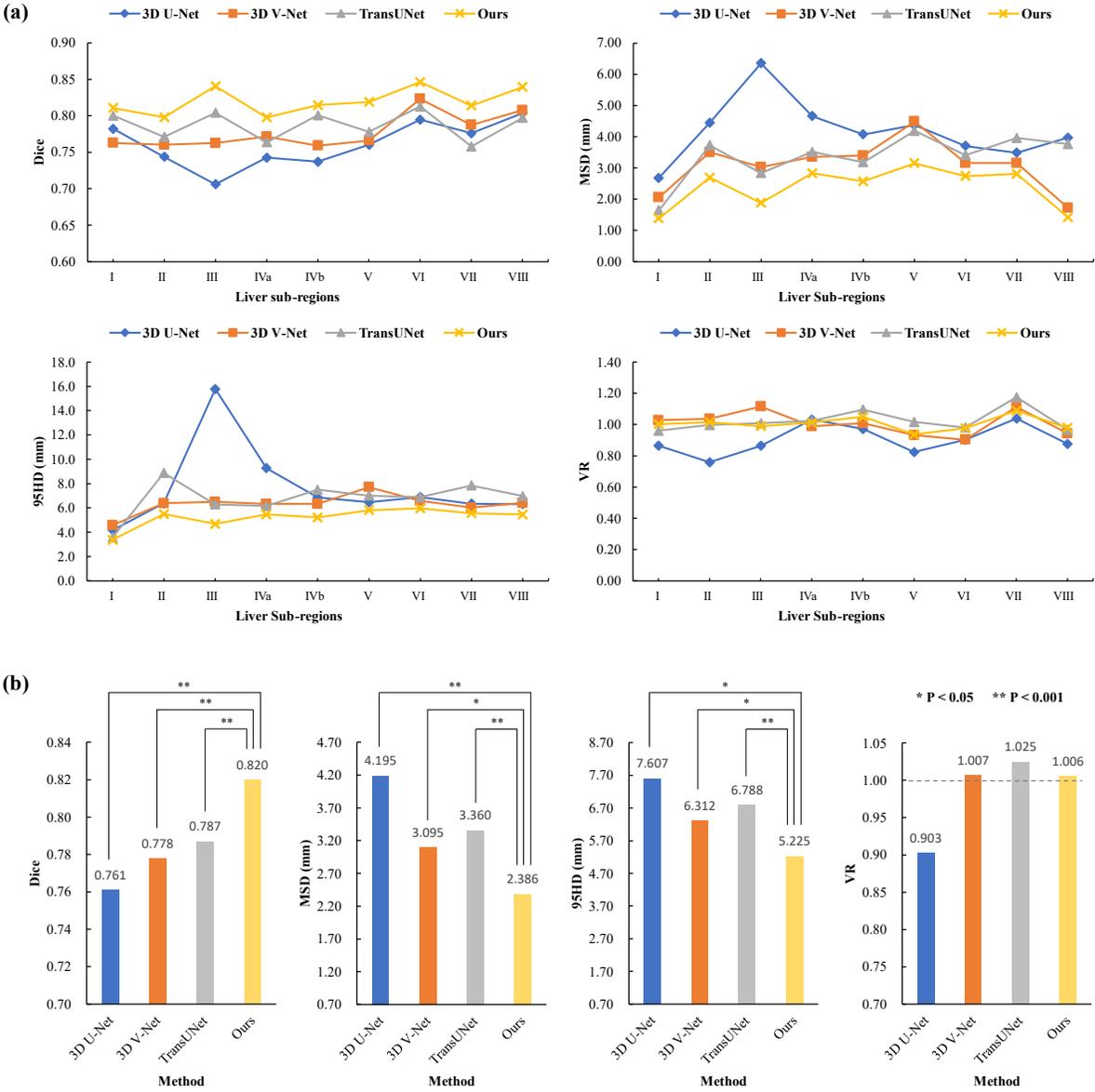

**Fig. 5 Quantitative comparison of various methods for Couinaud segmentation.** (a) The performance of various models, including 3D U-Net, 3D V-Net, TransUNet, and our Liverformer, is assessed using Dice, MSD, 95HD, and VR metrics across all liver sub-regions (I–VIII). (b) The average metrics for each method are presented alongside the corresponding p-values for Dice, MSD, and 95HD, providing a comprehensive comparison.

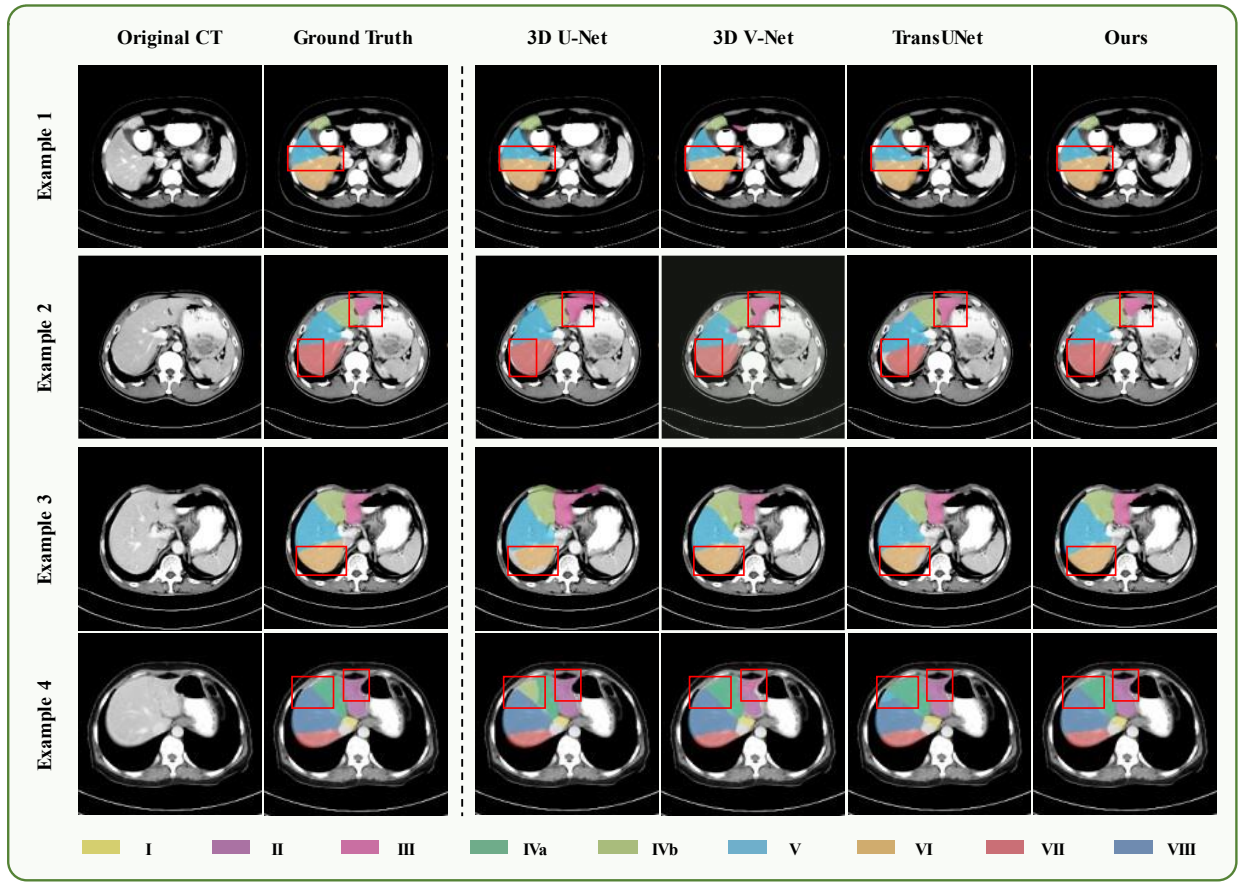

**Fig. 6 Qualitative comparison with the state-of-the-art methods.** The first two columns show the original CT images and their corresponding ground truth segmentations. The third through seventh columns display the automated Couinaud segmentation results using 3D U-Net, 3D V-Net, TransUNet, and our proposed method, respectively. Each column features a representative example from a specific patient. Red boxes are used to highlight discrepancies in the segmentation results, demonstrating the superior performance of our method.

To intuitively visualize segmentation performance, we present the Dice results in Fig. 5, along with additional critical metrics, including MSD, 95HD, and VR for a comprehensive evaluation. Fig. 5(a) showcases that our approach consistently surpasses the performance of other methods across all hepatic segments (I–VIII) in terms of Dice, MSD and 95HD, with VR also showing favorable outcomes. Additionally, the average results of 2.386±1.123 mm for MSD, 5.225±1.717 mm for 95HD, and 1.006±0.170 for VR, as shown in Fig. 5(b), indicate high boundary precision and volumetric consistency.

All p-values are below 0.05, with several even falling under 0.001, indicating that the improvements achieved by our approach are statistically significant or highly significant against other techniques. Notably, the segmentation accuracy in segments III and VIII, which are often challenging due to their anatomical positions and variability, highlights the robustness of our approach. Overall, this comprehensive evaluation and analysis validate the practical applicability of our method in real-world medical imaging scenarios.

**Qualitative Analysis.** We performed a qualitative evaluation of the automatic segmentation results for Couinaud segments, using representative examples from different patients for better visualization, as shown in Fig 6. The original CT images and their respective ground truth segmentations are shown in the first two columns, serving as references for evaluating the accuracy of the segmentation algorithms. The third to sixth columns present the segmentation outputs from 3D U-Net, 3D V-Net, TransUNet, and our proposed LiverFormer, respectively. As observed, 3D U-Net and 3D V-Net often struggle with maintaining accurate boundaries, especially in complex regions with high variability, leading to segmentation leaks or gaps, such as Segment III in Example 2, Segment VI in Example 3, and Segment III and IV in Example 4. TransUNet generally shows improved boundary adherence and internal consistency compared to 3D U-Net and 3D V-Net, but still exhibits occasional inaccuracies and misclassifications in complex regions, such as Segment VII in Example 2 and Segment VIII in Example 4. In contrast, our proposed LiverFormer excels in providing more precise boundaries, more accurately following the contours of Couinaud segments even in challenging regions. This demonstrates its ability to capture fine-grained details and complex anatomical structures. LiverFormer achieves high internal consistency within Couinaud segments, generating robust segmentation results that closely match the ground truth. It effectively segments small and difficult regions, accurately capturing subtle anatomical variations and ensuring comprehensive segmentation of Couinaud segments, unlike the other methods which exhibit boundary inaccuracies or inter-slice misclassification, such as Segments III and VII in Example 2, and Segments III, IV and VIII in Example 4. These qualitative results, alongside the quantitative metrics, validate the practical applicability and superior performance of LiverFormer in real-world medical imaging scenarios. Our method's ability to deliver accurate, detailed, and robust segmentation underscores its potential for enhancing clinical assessment and treatment planning.

## 4 Discussion and conclusion

Accurately delineating liver segments in CT and MRI images is crucial for clinical decision-making, especially in radiation treatment planning and surgical resection, as it can significantly improve long-term survival by reducing the risk of local recurrence. However, the task is challenging due to visual ambiguities between Couinaud segments, and manual segmentation, usually performed by radiologists, is laborious, time-intensive, and prone to inconsistencies both among and within observers. To overcome these difficulties, we propose a robust and precise deep learning approach for fully automating Couinaud segmentation. Although MRI offers superior soft tissue differentiation, allowing for clearer distinctions between various tissue types and simplifying the segmentation process, it is hindered by longer scan durations, higher costs, and potential accessibility limitations. Consequently, our focus is on Couinaud segmentation using CT images, which are more widely used due to their speed, resolution, and cost-effectiveness, despite the challenges posed by limited soft tissue contrast. In this study, we introduced LiverFormer, an innovative hybrid architecture combining 3D CNNs and Transformers, specifically designed for Couinaud segmentation. Our method achieved superior quantitative performance with an average Dice score of 0.802±0.062 without the use of data augmentation. This highlights the efficiency of combining fine-grained spatial details from CNNs with the broad contextual understanding offered by Transformers for feature extraction, surpassing advanced models like 3D U-Net, 3D V-Net, and TransUNet. Quantitative analysis using MSD, 95HD and VR further demonstrated the robustness of our network design in maintaining precise boundaries, high volumetric consistency, and effective handling of small and difficult segments. These results are consistent with the expanding research that emphasizes the promise of hybrid architectures in medical image segmentation. Considering the limited size and diversity of our medical imaging dataset due to data acquisition challenges and annotation complexity, we propose enhancing the robustness and generalizability of our deep learning model based on synthesis of realistic labeled examples through data augmentation. This resulted in higher performance with Dice score of 0.820±0.049. Moreover, the differences in Dice, MSD, and 95HD values between our model and the others were statistically evaluated using a paired t-test. As illustrated in Fig. 5, all p-values fell below 0.05, confirming that the enhancements offered by our method are statistically significant when compared to the other models.

While the outcomes are encouraging, our study is not without its limitations. Firstly, as a retrospective, single-institution study, the dataset may not fully capture the diversity encountered in clinical practice. Even with our effective data augmentation strategy for synthesizing realistic images, future work should focus on validating LiverFormer on larger, more diverse datasets collected prospectively from multiple centers to ensure generalizability. Secondly, our model's segmentation performance was validated only on CT images. Extending this validation to other imaging modalities, such as MRI and PET, would further explore the model's robustness and versatility. Additionally, the data augmentation was performed using the traditional deformable registration tool ANTs, which is time-consuming and somewhat lack of flexibility and adaptability. Developing a learning-based model capable of randomly applying logical and realistic transformations to capture nonlinear anatomical changes and variations in imaging intensity would be highly advantageous. This would allow for the rapid generation of a diverse set of realistic images from a limited pool of labeled data. Furthermore, our study excluded complex cases involving advanced cirrhosis, large tumors, partial hepatectomy, liver transplantation, or severe vascular problems. These conditions can significantly alter the liver's anatomy, potentially posing great challenges to our model. Future research should address these issues by involving those intricate yet common cases. By addressing these limitations, we can further enhance the generalizability, robustness, and applicability of LiverFormer in diverse clinical settings.

In conclusion, LiverFormer represents a significant advancement in Couinaud segmentation, showcasing the effectiveness of our hybrid 3D CNN-Transformer architecture and robust data augmentation strategies. Enhanced segmentation accuracy holds substantial clinical implications, facilitating more precise diagnosis for early detection and characterization of liver diseases, and improving treatment planning and outcomes. Furthermore, our method has the potential to greatly reduce the workload on radiologists, thereby improving workflow efficiency and allowing more time for critical decision-making and patient care. By setting a new benchmark in Couinaud segmentation, our work lays the foundation for future research and clinical applications, opening new avenues for the integration of advanced AI techniques in medical imaging and healthcare.